\documentclass[a4paper]{article}

\usepackage[english]{babel}
\usepackage[utf8x]{inputenc}
\usepackage{amsmath}
\usepackage{amssymb}
\usepackage{amsthm}
\usepackage{graphicx}
\usepackage[colorinlistoftodos]{todonotes}
\usepackage{cite}

\theoremstyle{plain}
\numberwithin{equation}{section}
\newtheorem{defi}{Definition}
\numberwithin{defi}{section}
\newtheorem{lemme}[defi]{Lemma}
\newtheorem{coro}[defi]{Corollary}
\newtheorem{prop}[defi]{Proposition}
\newtheorem{theo}[defi]{Theorem}
\theoremstyle{definition}
\newtheorem{example}[defi]{Example}
\theoremstyle{remark}
\newtheorem{remark}[defi]{Remark}

\def\nn{\mathbb{N}}
\def\zz{\mathbb{Z}}
\def\rr{\mathbb{R}}
\def\cc{\mathbb{C}}
\def\H{{\mathcal H}}
\def\h{{\mathfrak h}}
\def\C{{\mathcal C}}
\def\S{{\mathcal S}}
\def\pre{\noindent{\bf Proof: }}
\def\fin{$\Box$}
\def\tr{\mathrm{Tr}}
\def\id{\mathrm{Id}}

\makeatletter
\newcommand{\oset}[2]{%
  {\mathop{#2}\limits^{\vbox to -.5\ex@{\kern-\tw@\ex@
   \hbox{\scriptsize #1}\vss}}}}
\makeatother

\def\pp{\mathbb{P}}

\def\V{{\mathcal V}}
\def\rhoinv{\rho^{\mathrm{inv}}}
\def\eps{\varepsilon}
\def\red{\color{red}}

\def\connect{{\rightarrow}}

\def\biconnect{{\leftrightarrow}}

\def\enc{\mathrm{Enc}}

\def\ket#1{|#1\rangle}
\def\braket#1#2{\langle #1  , #2\rangle}
\def\ketbra#1#2{|#1\rangle\langle#2|}
\def\supp{\mathrm{supp}}

\title{Irreducible decompositions and stationary states of quantum channels}
\author{Raffaella Carbone\\Dipartimento di Matematica dell’Universit\'a di Pavia\\ via Ferrata, 1, 27100 Pavia, Italy\\\texttt{raffaella.carbone@unipv.it}\\ 
\\Yan Pautrat\\Laboratoire de Math\'ematiques\\Universit\'e Paris-Sud\\91405 Orsay Cedex, France\\\texttt{yan.pautrat@math.u-psud.fr}}
\begin{document}
\bibliographystyle{abbrv}
\maketitle

{\noindent\bf Abstract.} 
For a quantum channel (completely positive, trace-preserving map), we prove a generalization to the infinite dimensional case of a result by Baumgartner and Narnhofer (\hspace{1sp}\cite{BN}). This result is, in a probabilistic language, a decomposition of a general quantum channel into its irreducible positive recurrent components. This decomposition is related with a communication relation on the reference Hilbert space. This allows us to describe the full structure of invariant states of a quantum channel, and of their supports.

\section{Introduction}

The time-evolution of a closed quantum system is usually described as the conjugation by a group of unitary operators on the Hilbert space representing the state space of the system. When the system is open, that is, exchanges energy with its surroundings, the situation is more complicated and rigorous treatment usually requires approximations. The most standard approach was put on solid mathematical ground by Davies in the seventies (in \cite{Dav}, see also \cite{DF}), and leads to describe the system's evolution by a semigroup $(\Phi_t)_{t\in \rr_+}$ of linear maps on the set of states (\textit{i.e.} positive, normalized functionals acting on the set of operators on the Hilbert space) with specific algebraic properties (see section~\ref{section StatesQChannels}). Many features of these continuous parameter semigroups are already contained in the case of discrete semigroups $(\Phi^n)_{n\in \nn}$. In addition, the interest in the discrete case was renewed by quantum computation theory (where the maps $\Phi$ model quantum gates, see \cite{NieChu}) and by quantum repeated interaction systems (see \cite {BJM}). We therefore restrict ourseleves to the discrete case, and focus on the study of $\Phi=\Phi^1$, a linear map which is completely positive and trace-preserving. Such a map is called a quantum channel.

The study of ergodic properties of an open quantum system is therefore related to the study of invariants of $\Phi$, and of the associated spectrum. Analogies with operators associated with Markov chains (see Example \ref{example_MarkovChain}) inspired the development of a notion of irreducible quantum channel by various authors in the seventies and eighties (\hspace{1sp}\cite{AHK},  \cite{WE}, \cite{EHK}, \cite{Gro}), with different (and sometimes conflicting) definitions and implications. A vision of irreducibility as related to an intuitive notion of trajectories (as for Markov chains), however, was not developed explicitly before the work of Baumgartner and Narnhofer in \cite{BN}, where it is done in the case of a finite-dimensional Hilbert space. This vision allows to describe the decomposition of a reducible quantum channel into a sum of irreducible ones. In addition, a fine study of these decompositions leads to a description of the full structure of invariant states of a general quantum channel.

In \cite{CP1}, we studied open quantum random walks, a special class of evolutions belonging to the above case. This led us to restate and extend the results of \cite{BN} to the case of open quantum random walks, which required in particular an extension to the infinite dimensional case. Our proofs, however, apply to a wider class of evolutions than just quantum random walks. We therefore describe our results in full generality here.

The structure of this article is as follows. In section \ref{section StatesQChannels}, we describe our framework and in particular the evolutions $\Phi$ of interest, the so-called quantum channels. In section \ref{section_irreducibility}, we recall the different notions of irreducibility. In section~\ref{section_enclosures}, we define enclosures, our key tool, which originate in \cite{BN}. In section~\ref{section_invariantstates}, we describe the relation between enclosures and supports of invariant states. In section \ref{section_decomposition} we discuss the structure of invariant states of a simple reducible evolution. In section \ref{section_irreducibledecompositions}, we state our general decomposition theorem, that describes irreducible decompositions of evolutions and the general structure of the set of invariant states. In section \ref{section_examples}, we apply these results to a number of examples.

\paragraph{Acknowledgements.} RC gratefully acknowledges the support of PRIN project 2010MXMAJR 
and GNAMPA project ``Semigruppi markoviani su algebre non commutative'', and YP the support of ANR project n${}^\circ$ANR-14-CE25-0003. YP wishes to thank Julien Deschamps for enlightening discussions.

\section{States and Quantum Channels} \label{section StatesQChannels}

In this section we give a short summary of the theory of quantum channels, \textit{i.e.} completely positive, trace-preserving maps on an ideal of trace-class operators. We fix a separable Hilbert space $\mathcal H$, which is supposed to play the role of a state space for a quantum system. We denote by $\mathcal I_1(\H)$ the set of trace-class operators on $\H$ (see \cite{RS1}), and equip it with the topology induced by the trace norm. We recall that the topological dual $\mathcal I_1(\mathcal H)^*$ can be identified with the algebra $\mathcal B(\mathcal H)$ of bounded linear operators through the Schatten duality $(\rho,X)\mapsto \tr(\rho\, X)$. Therefore, the topology of $\mathcal I_1(\H)$ is the same as the weak topology induced by~$\mathcal B(\mathcal H)$.
We also recall that an operator $X$ on $\mathcal H$ is called nonnegative (respectively positive or positive definite), denoted $X\geq0$ (resp. $X>0$), if for $\varphi\in \mathcal H\setminus\{0\}$,  one has $\langle \varphi, X\, \varphi\rangle \geq 0$ (resp. $\langle\varphi, X\, \varphi\rangle>0$).

The states of a system will be represented by an operator belonging to a specific class: 
\begin{defi} \label{defi_state}
An operator $\rho$ is called a state if it is self adjoint (\textit{i.e.} $\rho=\rho^*$), nonnegative, and is trace-class with trace one. We denote by $\S(\H)$ the set of states on $\H$. A state is called faithful if it is positive definite.
\end{defi}

\begin{remark} \label{remark_normalornot}
In the literature, a state is sometimes defined as a positive linear form on $\mathcal B(\mathcal H)$ mapping $\id$ to $1$, \textit{i.e.} as an element of the set
\[\mathcal B(\H)^*_{+,1}= \{ \eta\in B(\H)^* \ \mbox{s.t.}\ \eta(X)\geq 0 \ \mbox{for}\ X\geq 0 \ \mbox{and}\ \eta(\id)=1\}\]
equipped with the weak-* topology. The objects defined in Definition \ref{defi_state} are then called normal states. Obviously $\mathcal S(\H)$ is homeomorphic to a subset of~$\mathcal B(\H)^*_{+,1}$.
\end{remark}

 Consider now a linear map $\Phi$ on $\mathcal I_1(\H)$. We say that this map is positive if it maps nonnegative elements of $\mathcal I_1(\H)$ to nonnegative elements of $\mathcal I_1(\H)$. We say that it is $n$-positive, for $n\in \nn$, if the map $\Phi\otimes \id_{\mathcal M_n(\cc)}$ is positive as a map on $\mathcal I_1(\H\otimes \cc^n)$;  and completely positive if it is $n$-positive for any $n$ in $\nn$. We say that it is trace-preserving if, for any $\rho\in \mathcal I_1(\H)$, one has $\tr(\Phi(\rho))=\tr(\rho)$; in particular a positive trace-preserving map induces a map on $\S(\H)$. Our main objects of interest will be maps that are completely positive and trace-preserving:
 \begin{defi}
 A completely positive, trace-preserving map on a space $\mathcal I_1(\H)$ is called a quantum channel on $\H$.
 \end{defi}
 
 \begin{remark}
 A positive linear map on $\mathcal I_1(\H)$ is automatically bounded (see Lemma 2.2 in \cite{Sch}), so that it is weak-continuous.
 \end{remark}
 
 The following theorem states a well-known fact about quantum channels (see \cite{Kraus}, \cite{NieChu}):
 \begin{theo}
 A linear map $\Phi$ on $\mathcal I_1(\H)$ is completely positive if and only if there exists a family $(V_i)_{i\in I}$ of operators on $\H$ such that for any $\rho$ in $\mathcal I_1(\H)$,
 \begin{equation}\label{eq_KrausForm} 
 \Phi(\rho)=\sum_{i\in I} V_i \rho V_i^*.
 \end{equation}
If in addition $\Phi$ is trace-preserving, then the operators $V_i$ satisfy the relation
\[\sum_{i\in I}  V_i^* V_i = \id_\H.\]
 \end{theo}
The decomposition \eqref{eq_KrausForm} is called a Kraus form of $\Phi$, and the family $(V_i)_{i\in I}$ an unravelling. Note that an unravelling of $\Phi$ is not unique (see \cite{NieChu} for more details).

We have mentioned that a source of inspiration is the analogy between quantum channels and Markov chains. In the following example we point out that Markov chains are a special case of quantum channels. Note that, for any two vectors~$x$ and $y$ in a Hilbert space $\H$ with scalar product $\braket{\cdot}{\cdot}$ (which we assume is antilinear in the left variable), we denote by~$\ketbra xy$ the map $z \mapsto \braket y z\ x$.

\begin{example} \label{example_MarkovChain}
  Consider a Markov chain $(X_n)_n$ on a countable set $E$ with transitions $p_{i,j}=\pp(X_{n+1}=i\,|\,X_n=j)$. If we let $\H$ be $\ell^2(E)$, the set of (complex valued) square-summable sequences indexed by $E$, denote by $(e_i)_{i\in E}$ the canonical orthonormal basis, and consider $V_{i,j} =\sqrt{p_{i,j}\vphantom{1}}\, \ketbra{e_i}{e_j}$ for $i,j$ in~$E$, then \eqref{eq_KrausForm} defines a quantum channel, and any invariant state of $\Phi$ is of the form $\rho=\sum_{i\in E} \pi_i \ketbra{e_i}{e_i}$ with $(\pi_i)_{i\in E}$ an invariant probability measure for the Markov chain.
\end{example}

\begin{remark}\label{remark_TPnormone}
Trace-preservation of a map $\Phi$ is equivalent to $\Phi^*(\id)=\id$. The adjoint $\Phi^*$ is then a positive, unital (i.e. $\Phi^*(\id)=\id$) map on $\mathcal B (\mathcal H)$, and by the Russo-Dye theorem (\hspace{1sp}\cite{RD}) one has $\|\Phi^*\|=\|\Phi^*(\id)\|$ so that $\|\Phi\|=\|\Phi^*\|=1$.
\end{remark}

A quantum channel represents the (discrete) dynamics of an open quantum system in the Schrödinger picture (see \cite{NieChu} for more details). We denote by~$\mathcal F(\Phi)$ the subset of $\mathcal I_1(\H)$ of invariant elements of $\Phi$ and we will be specifically interested in the set $\S(\H)\cap \mathcal F(\Phi)$ of invariant states, \textit{i.e.} elements of $\S(\H)$ that are invariant by $\Phi$. 

For $\rho$ a state we will consider its support, which is defined as the range of the projection $\id-P_0(\rho)$, where
\[P_0(\rho)=\sup\{P\ \mathrm{orthogonal\ projection\ s.t.}\, \rho(P)=0\}.\]
The supremum taken above is considered with respect to the order induced by the relation $\geq$ for operators, and always exists in the present situation. Following \cite{FV}, we denote: 
\[ \mathcal R = \mathrm{sup}\{\mathrm{supp}\,\rho\, |\, \rho \mbox{ an invariant state}\} \]
so that by definition, $\mathrm{supp}\, \rho\subset \mathcal R$ if $\rho$ is an invariant state. This space is often called the fast recurrent space, in parallel with the classical case, where the fast recurrent configurations are the ones which support the invariant probability laws. The orthogonal of $\mathcal R$ is
\[\mathcal D = \{x \in \mathcal H \, |\, \langle x,\rho\,x\rangle =0 \ \mbox{ for any invariant state }\rho \}.\] 

\begin{remark}
In \cite{BN}, the states $\mathcal R$ and $\mathcal D$ are defined without reference to the set of invariant states, as
\[\mathcal D = \{x\in \mathcal H\, |\,\langle x,\Phi^n(\rho)\, x\rangle\underset{n\to\infty}{\longrightarrow}0 \mbox{ for any state }\rho\}\]
and $\mathcal R=\mathcal D^\perp$. These different definitions of $\mathcal R$ and $\mathcal D$ are equivalent in finite dimension.
\end{remark}

\begin{remark}
The space $\mathcal D$ is the sum of the transient and slow recurrent subspaces, as defined in \cite{Um}. 
\end{remark}

\section{Irreducibility}\label{section_irreducibility}

Before we discuss decompositions of quantum channels, we need to discuss the relevant reducing components of the decomposition, \textit{i.e.} irreducible quantum channels. As we will see in Proposition \ref{prop_Schrader} and Remark \ref{remark_remarksirreducibility}, irreducibility is strongly connected with the uniqueness of the invariant state.

As we already mentioned in the introduction, however, different definitions of irreducibility of quantum channels can be found in the literature. We will briefly recall them here. First we need to define some relevant concepts:
\begin{defi}\label{def-irreducibility}
Let $\Phi$ be a quantum channel on $\S(\H)$. We say that an orthogonal projection $P$:
\begin{itemize}
\item reduces $\Phi$ if we have $\Phi\big(P\mathcal I_1(\mathcal H)P\big) \subset P\mathcal I_1(\mathcal H)P$,
\item is subharmonic for $\Phi^*$ if $\Phi^*(P)\geq P$.
\end{itemize}
\end{defi}

The complete proof of the following Proposition is given in \cite{CP1}:
\begin{prop} \label{prop_defirreducibility}
Let $\Phi$ be a quantum channel on $\mathcal I_1(\H)$. The following properties are equivalent:
\begin{itemize}
\item $\Phi$ is Davies-irreducible: the only orthogonal projections reducing $\Phi$ are $P=0$ and $\id$;
\item the only orthogonal projections that are subharmonic for $\Phi^*$ are $P=0$ and~$\id$;
\item ergodicity: for any state $\rho$, the operator $(\exp t\Phi) (\rho)$ is definite-positive for any $t>0$.
\end{itemize}
\end{prop}
We say that $\Phi$ is irreducible if and only if any of the properties in Proposition~\ref{prop_defirreducibility} holds.

\begin{remark}  \label{remark_remarksirreducibility}
Regarding the above concepts and their interrelations:
\begin{enumerate}
\item the equivalence between the first two properties follows from the simple observation that an orthogonal projection reduces $\Phi$ if and only if it is subharmonic for $\Phi^*$  (see \cite[Proposition 3.3]{CP1});
\item the definition of ergodicity given here originates in \cite{Sch}, and extends the definition given in \cite{EHK} to infinite-dimensional $\H$;
\item there exists yet another notion of irreducibility: one says that $\Phi$ is Evans-irreducible if the only orthogonal projections that are harmonic for $\Phi$, \textit{i.e.} such that $\Phi^*(P)=P$, are $P=0$ and $\id$. Clearly Davies-irreducibility implies Evans-irreducibility, but the converse is not true in general.
\end{enumerate}
\end{remark}

In the same fashion as for Markov semigroups, there exists a Perron-Frobenius theorem related to the property of irreducibility. We state it in the next proposition, in a form essentially due to Schrader in \cite{Sch}:
\begin{prop}\label{prop_Schrader}
Let $\Phi$ be a quantum channel on $\mathcal I_1(\mathcal H)$, and assume it has an eigenvalue $\lambda$ of modulus $1$, 
with eigenvector $\rho$. Then:
\begin{itemize}
\item $1$ is also an eigenvalue, with eigenvector $|\rho|=(\rho^*\rho)^{1/2}$,
\item if $\Phi$ is irreducible, then $\lambda$ is a simple eigenvalue and $|\rho|>0$.
\end{itemize}
\end{prop}
\begin{remark} \label{remark_PFHeisenberg}
Proposition \ref{prop_Schrader} still holds if $\Phi$ is not completely positive and trace-invariant, but simply 2-positive and trace-invariant. For this reason, the same statement holds when the map $\Phi$ on $\mathcal I_1(\mathcal H)$ is replaced with the map $\Phi^*$ on $\mathcal B(\mathcal H)$, and all subsequent results about quantum channels will hold for 2-positive and trace-invariant maps on $\mathcal I_1(\H)$, as long as they do not involve the Kraus form or unravelling of $\Phi$.  
\end{remark}

An immediate consequence of this proposition is that an irreducible quantum channel on $\mathcal I_1(\H)$ has at most one invariant state. In sections~\ref{section_decomposition} and \ref{section_irreducibledecompositions} we will study the relations between the invariant states of a reducible quantum channel and the invariant states of its irreducible components.
 
\section{Enclosures and communicating classes} \label{section_enclosures}

For Markov chains, it is well-known that irreducibility is related with the notion of communication within the induced graph. In addition, communicating classes have an explicit description as orbits of points, and are the relevant objects to break down a reducible Markov chain into irreducible ones. In this section we introduce the notion of enclosure, that will parallel the notion of closed set for Markov chains, and allow us to study irreducible decompositions of quantum channels.

\begin{defi} \label{defi_enclosures}
 Let $\Phi$ be a quantum channel. A closed subspace $\V$ is an enclosure for $\Phi$ if, for any state $\rho$, $\supp\, \rho \subset \V$ implies $\supp\,\Phi(\rho)\subset\V$.
\end{defi}

We will call nontrivial any enclosure which is neither $\{0\}$ nor $\H$. Clearly, a subspace $\V$ is an enclosure if and only if it is the range of a reducing orthogonal projector. Therefore, a quantum channel $\Phi$ is irreducible if and only if it has no nontrivial enclosures. This shows that enclosures are relevant to the notion of irreducibility. 
\smallskip

We now prove a simpler characterization of enclosures:
\begin{lemme} \label{lemme_enclosure}
A vector subspace $\V$ of $\H$ is an enclosure if and only if, for any~$x$ in $\V$ with $\|x\|=1$, the state $\Phi(\ketbra xx)$ has support in $\V$.
\end{lemme}
\pre

Let $\rho$ be a state with support in $\V$. The spectral decomposition of $\rho$ is of the form $\sum_{i\in I} \lambda_i \ketbra{e_i}{e_i}$ with $\lambda_i>0$, $\sum_{i\in I} \lambda_i=1$ and $e_i\in \V$. Therefore, $\supp \, \Phi(\ketbra{e_i}{e_i})\subset \supp\,\Phi(\rho)$, which shows the direct implication; in addition, the support of $\Phi(\rho)$ is the supremum of the projectors on the ranges of $\Phi(\ketbra{e_i}{e_i})$ and this shows the converse.
\fin

This has the following useful corollary. Note that, for $(\V_i)_{i\in I}$ a family of closed subspaces of $\H$, we denote by \textit{e.g.} $\V_1+\V_2+\ldots$ or $\sum_{i\in I} \V_i$ the closed vector space generated by $\bigcup_{i\in I}\V_i$.
\begin{coro} \label{coro_sumsenclosures}
Let $\V_1$ and $\V_2$ be two enclosures. The closed subspace $\V_1+\V_2$ is also an enclosure.
\end{coro}
\pre
By a direct computation, $\ketbra {x_1+x_2}{x_1+x_2}\leq 2\,\ketbra{x_1}{x_1}+2\,\ketbra{x_2}{x_2}$ for $x_1,x_2$ in $\V_1,\V_2$ respectively. Applying Lemma \ref{lemme_enclosure} shows that $\V_1+\V_2$ is an enclosure.
\fin
\smallskip

This allows us to obtain an explicit characterization of enclosures in terms of unravellings of $\Phi$, and connect them to a notion of orbit under the action of possible transitions of $\Phi$.
\begin{prop}\label{prop_enclosures}
Consider a quantum channel $\Phi$ with unravelling $(V_i)_{i\in I}$. A subspace $\V$ of $\H$ is an enclosure if and only if $V_i\, \V \subset \V$ for any $i$.
\end{prop}

\pre
The proposition follows from Lemma \ref{lemme_enclosure} and the fact that, by the trace norm continuity of $\Phi$ one has for any $x \in\V$,
\begin{equation}\label{eq_pippo}
\Phi(\ketbra xx) = \sum_{i \in \, I}  \ketbra{V_ix}{V_ix}. \quad\Box
\end{equation}
%


Our goal is to consider enclosures defined as the set of points accessible from a given initial $x\in \H$. Proposition \ref{prop_enclosures} suggests a natural definition.
\begin{prop}\label{prop_enclosures2}
Let $\Phi$ be a quantum channel on $\mathcal I_1(\H)$. Let $(V_i)_{i\in I}$ be an unravelling of $\Phi$. For $x$ in $\mathcal H\setminus\{0\}$, we call enclosure generated by $x$ the closed vector space 
\begin{equation} \label{eq_EnclosureunRavelling}
\enc(x)=\cc x\,+\,\overline{\mbox{\rm span} \{V_{i_1}\cdots V_{i_n}\, x, \,|\, n\in \nn^*,\,i_1,.. i_n\in I \}}.
\end{equation}
With this definition, the space $\enc(x)$ is the smallest enclosure containing $x$.
\end{prop}

%
%
\pre
It follows from \eqref{eq_pippo} that definition \eqref{eq_EnclosureunRavelling} also satisfies
\begin{equation} \label{eq_EnclosureEqDefinition}
\enc(x)=\overline{\mathrm{span}\{\mathrm{supp}\, \Phi^n(\ketbra xx),\, n\ge 0\}}.
\end{equation}
This shows that definition  \eqref{eq_EnclosureunRavelling} is independent of the choice of unravelling. The fact that $\enc(x)$ is an enclosure then follows from Proposition \ref{prop_enclosures}.
%
\fin
\begin{remark}
This implies in particular that a quantum channel $\Phi$ is irreducible if and only if $\mathcal H=\enc(x)$ for any $x$ in $\mathcal H\setminus\{0\}$.
\end{remark}

We can define a notion of accessibility among vectors in $\H$, related to the notion of enclosure, and consider an equivalence relation. We will argue, however, that this will not immediately provide us with an interesting decomposition of a quantum channel.
\begin{defi}
For $x$, $y$ in $\mathcal H$, we say that:
\begin{itemize}
\item $y$ is accessible from $x$ (and denote it by $x \connect y$) if $y\in\enc(x)$;
\item $y$ and $x$ communicate (and denote it by $x \biconnect y$) if~$\enc(x)~=~\enc(y)$.
\end{itemize}
\end{defi}

One can immediately observe that accessibility is a transitive relation, and communication is an equivalence relation. We denote by ${\mathcal C}(x)$ the equivalence class of a vector $x$ in $\mathcal H$ for the relation $\biconnect$,
$$
\C(x) = \{ y\in \enc(x)\ \mathrm{ s.t. }\ x\in\enc(y)\}.
$$
An equivalence class of a vector $x $ by $\biconnect$ is a subset of $\enc (x)$ but it is not a vector space since, for $x\neq 0$, $\C(x)$ cannot contain $0$. Even adding the point $0$ may fail to make $\C(x)$ a vector space, as the next example shows.

\begin{example} \label{example_23}
Take $\H=\cc^2$ and denote by $e_1,e_2$ its canonical basis. Consider a quantum channel $\Phi$ on $\mathcal I_1(\H)$ with unravelling $(V_1,V_2)$ given by
$
V_1=\sqrt p \left(\!\begin{smallmatrix}
0 & 1 \\ 0 & 0
\end{smallmatrix}\!\right)$ and
$ V_2=\left(\!\begin{smallmatrix}
1 & 0 \\ 0 & \sqrt{1-p}
\end{smallmatrix}\!\right)$
for some $p\in(0,1)$
so that, for $\rho=\left(\!\begin{smallmatrix}
\rho_{1,1} & \rho_{1,2} \\ \rho_{2,1} & \rho_{2,2}\end{smallmatrix}\!\right)$ in $\mathcal I_1(\H)$, we have
$$
\Phi(\rho)=
\begin{pmatrix}
p\rho_{2,2}+\rho_{1,1} & \sqrt{1-p}\; \rho_{1,2} \\ \sqrt{1-p}\; \rho_{2,1} & (1-p)\rho_{2,2}
\end{pmatrix}.
$$
By an immediate direct computation, the state $\ketbra{e_1}{e_1}$ is the only invariant state of this map.
We want to describe the equivalence classes and the enclosures of the map $\Phi$.

We notice that, for any vector $u={}^t(u_1,u_2)$ in $\cc^2$,
$$
\ketbra uu=
\begin{pmatrix}
|u_1|^2 & u_1 \bar u_2 \\ \bar u_1 u_2 & |u_2|^2
\end{pmatrix}
\  \mbox{so that}\ 
\Phi(\ketbra uu)=
\begin{pmatrix}
p|u_2|^2+|u_1|^2& \sqrt{1-p}\; u_1 \bar u_2 \\ \sqrt{1-p}\; \bar u_1 u_2 & (1-p)|u_2|^2
\end{pmatrix}.
$$ 
It is immediate that $\Phi(\ketbra uu)$ is a positive definite matrix whenever $u_2\neq 0$, so that
\begin{itemize}
\item $\mathrm {supp} \,\Phi^n(\ketbra {e_1}{e_1})=\cc\,e_1$ for all $n\ge 0$,
\item for $u_2\neq 0$, $\mathrm {supp} \,\Phi^n(\ketbra uu)=\cc^2$ for all $n\ge 1$.
\end{itemize}
Identity \eqref{eq_EnclosureEqDefinition} allows us to  determine all the enclosures and equivalence classes:
\begin{itemize}
\item  $\enc(0)=\C(0)=\{0\}$,
\item $\enc(e_1)=\cc\,e_1$ and $\C(e_1)=\enc(e_1)\setminus\{0\}$,
\item for all $u\in\cc^2\setminus \cc\,e_1$, $\enc(u)=\cc^2$ and $\C(u)=\cc^2\setminus  \mathrm \cc\, e_1$.
\end{itemize}
\end{example}
\smallskip

Supports of invariant states, on the other hand, are always vector spaces. Therefore, the naive approach of considering the partition of $\H$ induced by the relation $\biconnect$ to obtain a relevant decomposition of a quantum channel into irreducible such maps fails, as it does not seem to involve the vector space structure. A natural idea, derived from the study of Markov chains, is to consider specifically minimal objects. We therefore give the following definition of a minimal enclosure:
\begin{defi}
Let $\V$ be an enclosure. We say that $\V$ is a minimal enclosure if any enclosure $\V'$ satisfying $\V'\subset \V$ is either $\{0\}$ or $\V$. We say that $\V$ is a minimal nontrivial enclosure if in addition $\V\neq\{0\}$.
\end{defi}
The following easy proposition shows that this notion is indeed relevant:
\begin{prop}
$\C(x)=\enc(x)\setminus \{0\}$ if and only if $\enc(x)$ is a minimal nontrivial enclosure.
\end{prop}

\pre
If $\C(x)=\enc(x)\setminus \{0\}$, then, for all $y$ in $\enc(x)\setminus \{0\}$, we have $\enc(x)=\enc(y)$ and consequently $\enc(x)$ is minimal.  Conversely, if $\V=\enc(x)$ is a minimal enclosure, for any $y$ in~$\V\setminus\{0\}$, $\enc(y)$ is a nontrivial enclosure contained in $\V$ so that $\enc(y)=\V$. Therefore $x\biconnect y$ and $\V=\C(x)$. 
\fin
%

\section{Enclosures and invariant states} \label{section_invariantstates}

Baumgartner and Narnhofer (in \cite{BN}) studied a decomposition of a quantum channel related to the supports of extremal invariant states, in the case of a finite dimensional space $\H$. In the present paper, we extend this analysis to the infinite dimensional case. For this we will need to relate extremal invariant states to minimal enclosures. We will see that the form of invariant states for the quantum channel is dictated by the uniqueness or non-uniqueness of the decompositions into minimal enclosures and that this is related to the existence of mutually non-orthogonal minimal enclosures. 

The first result is:
\begin{prop}\label{prop_SuppInvSt}
Let $\Phi$ be a quantum channel on $\H$.
\begin{enumerate}
\item The support of an invariant state is an enclosure.
\item The fast recurrent subspace $\mathcal R$ is an enclosure.
\end{enumerate}
\end{prop}

\pre
To prove the first point, fix an invariant state $\rho_0$, and let $\rho$ be another state with support contained in $\mathrm{supp}\,\rho_0$. Fix an orthonormal family of eigenvectors for $\rho_0$ generating $\mathrm{supp}\,\rho_0$, and let $X_0$ be the set of finite linear combinations of these vectors. This set $X_0$ is dense in $\mathrm{supp}\,\rho_0$ and for every $x$ in $X_0$ there exists $\lambda$ such that $\ketbra xx \leq \lambda \rho_0$.  Therefore there exists an approximation of~$\rho$ in the $\mathcal I_1(\mathcal V)$ norm sense by an increasing sequence of finite-dimensional operators $(\rho_p)_p$ 
such that for every $p$ there exists a $\lambda_p$ with $\rho_p\leq \lambda_p\rho_0$, so that $\Phi(\rho_p)\leq \lambda_p \Phi(\rho_0)$ and therefore $\mathrm{supp}\,{\Phi}(\rho_p)\subset\mathrm{supp}\,\rho_0$. The sequence $\Phi(\rho_p)$ is increasing and weakly convergent to $\Phi(\rho)$ so that $\mathrm{supp}\,\Phi(\rho)\subset\mathrm{supp}\, \rho_0$, which proves that $\mathrm{supp}\, \rho_0$ is an enclosure.

To prove the second point, associate with every invariant state $\rho$ the orthogonal projector $P_\rho$ on its support. Then the orthogonal projector $P$ on $\V$ is the supremum of the family $(P_\rho)_{\rho}$. By Proposition \ref{prop_SuppInvSt}, every $P_\rho$ is subharmonic, \textit{i.e.} $\Phi^*(P_\rho)\ge P_\rho$ for any invariant state $\rho$. Moreover, $\Phi^*(P)\ge \Phi^*(P_\rho)\ge P_\rho$ for any invariant $\rho$, so that $\Phi^*(P)\ge P$ and the conclusion follows. \fin

\begin{remark} \label{remark_Umanita}
The first point of the previous proposition has already been proven in \cite{FR} and \cite{Um} in the dual setting, {\it i.e.} considering reducing projections for $\Phi^*$. If $\H$ is separable, the second point can also be derived from a result from \cite{Um} which proves that there exists an invariant state with support equal to $\mathcal R$.
\end{remark} 
\begin{remark}
The converse of point 1 of Proposition \ref{prop_SuppInvSt} is not true. Consider Example \ref{example_MarkovChain} associated with the symmetric random walk on $\zz$. Then $\H=\ell^2(\zz)$ is an enclosure but the quantum channel $\Phi$ has no invariant state.
\end{remark}

\begin{prop}\label{prop_coherence}
Let $\mathcal V$ be an enclosure, $\mathcal W$ be a subspace of $\mathcal H$ which is in direct sum with $\mathcal V$, 
and $P_{\mathcal V}$ and $P_{\mathcal W}$ be the respective orthogonal projections.
Consider a state $\rho$ with support in $\mathcal V \oplus \mathcal W$ and denote
\[\rho_{\mathcal V}= P_{\mathcal V}\, \rho \,P_{\mathcal V},\quad \rho_{\mathcal W}= P_{\mathcal W}\, \rho \,P_{\mathcal W}\, \quad \rho_{\mathcal C}= P_{\mathcal V}\, \rho\, P_{\mathcal W}, \quad \rho_{\mathcal C}'=P_{\mathcal W}\,\rho \,P_{\mathcal V}; \]
similarly, decompose $\Phi(\rho)$ into $\Phi(\rho)_{\mathcal V}+\Phi(\rho)_{\mathcal W}+\Phi(\rho)_{\mathcal C}+\Phi(\rho)_{\mathcal C}'$.
Then
\begin{enumerate}
\item $P_{\mathcal W}\,( \Phi (\rho_C)+ \Phi (\rho'_C))\, P_{\mathcal W}=0$;
\item if $\mathcal Z$ is another enclosure with $\V\subset \mathcal Z\subset \mathcal R$, then $\mathcal Z \cap \mathcal V^\perp$ is an enclosure;
\item if $\mathcal W$ is also an enclosure, then
\[ 
\Phi (\rho)_{\mathcal V} =\Phi(\rho_{\mathcal V}) \quad \Phi (\rho)_{\mathcal W} =\Phi(\rho_{\mathcal W})
\quad \Phi (\rho)_{\mathcal C} =\Phi(\rho_{\mathcal C})
\quad \Phi (\rho)_{\mathcal C}' =\Phi(\rho'_{\mathcal C}).
\]
\end{enumerate}
\end{prop}

\pre

\begin{enumerate}
\item Let
$\kappa_{\pm \eps}=\frac1\eps \,\rho_{\mathcal V}\pm \,\rho_{\mathcal C}+\eps \, \rho_{\mathcal W}.$
We have $\kappa_{\pm\eps}\geq 0$ (as can be checked from $\langle u, \kappa_{\pm\eps}\, u \rangle = \langle u_{\pm\eps}, \rho\, u_{\pm\eps} \rangle$, where $ u_{\pm\eps}=\frac1{\sqrt \eps}\,P_{\mathcal V}u + \sqrt \eps\, P_{\mathcal W}u$), so that $\Phi(\kappa_{\pm\eps})\geq 0$, and, because $\mathcal V$ is an enclosure, the support of $\Phi(\rho_{\mathcal V})$ is contained in $\mathcal V$, so that
\[ P_{\mathcal W}\,\Phi(\kappa_{\pm \eps})\, P_{\mathcal W} = \pm P_{\mathcal W}\big(\Phi(\rho_{\mathcal C})+\Phi(\rho_{\mathcal C}')\big)P_{\mathcal W}+ \eps \,P_{\mathcal W}\, \Phi (\rho_{\mathcal W}) \, P_{\mathcal W}\geq 0,\]
and by necessity $ P_{\mathcal W}\,(\Phi(\rho_{\mathcal C})+\Phi(\rho'_{\mathcal C}))\,P_{\mathcal W}=0$. 
\item Consider $\mathcal W=\mathcal Z\cap\mathcal V^\perp$ and $\rho$ any invariant state; then 
\[
\rho_{\mathcal V}+\rho_{\mathcal W}+\rho_{\mathcal C} +\rho_{\mathcal C}' 
= \Phi(\rho_{\mathcal V}) + \Phi(\rho_{\mathcal W})+\Phi(\rho_{\mathcal C})+\Phi(\rho_{\mathcal C}').
\]
Projecting by $P_{\mathcal W}$ this yields $\rho_{\mathcal W}= P_{\mathcal W} \Phi(\rho_{\mathcal W})P_{\mathcal W} $, so that $P_{\mathcal V}\,\Phi(\rho_{\mathcal W})\, P_{\mathcal V}$ is positive with zero trace. Therefore $P_{\mathcal V}\,\Phi(\rho_{\mathcal W})\, P_{\mathcal V}=0$ which implies $P_{\mathcal V}\,\Phi(\rho_{\mathcal W})=\Phi(\rho_{\mathcal W})\, P_{\mathcal V}=0$ and so $\rho_{\mathcal W}=\Phi(\rho_{\mathcal W})$. 
As the support of a stationary state, $\mathrm{supp}\,\rho_{\mathcal W}= \mathrm{supp}\,\rho\cap\mathcal Z\cap\mathcal V^\perp$ is an enclosure. By point 2 of Proposition \ref{prop_SuppInvSt}, taking the supremum over all possible invariant states~$\rho$ we deduce that $\mathcal Z \cap \mathcal V^\perp$ is also an enclosure.
\item If $\mathcal V$ and $\mathcal W$ are enclosures, then $\mathrm{supp}\,\Phi(\rho_{\mathcal V})\subset \mathcal V$ and $\mathrm{supp}\,\Phi(\rho_{\mathcal W})~\subset~\mathcal W$. The conclusion follows from the previous points. $\Box$
\end{enumerate}

We will now discuss the connection between minimal enclosures and extremal invariant states, \textit{i.e.} states $\rho$ such that $ \rho = t\,\rho_1 + (1-t)\, \rho_2$,
with $\rho_1$, $\rho_2$ in $\mathcal S(\H) \cap \mathcal F(\Phi)$ and $t\in(0,1)$, implies $\rho_1=\rho_2=\rho$.
\begin{remark}
The distinction between states and normal states mentioned in Remark \ref{remark_normalornot} does not lead to an ambiguity: by Example 4.1.35 in \cite{BR1}, the set $\mathcal S(\H)$, when viewed as a subspace of $\mathcal B(\H)^*_{+,1}$, is a face, so that $\rho \in \mathcal S(\H)$ is extremal regarding convex decompositions in $\mathcal S(\H)\cap \mathcal F(\Phi)$ if and only if it is extremal regarding convex decompositions in $\mathcal B(\H)^*_{+,1} \cap \mathcal F(\Phi)$.
\end{remark}

\begin{coro} \label{coro_subinvariantstate}
For any enclosure $\V$ contained in $\mathcal R$, there exists an invariant state $\rho$ such that $\supp\, \rho \subset \V$.
\end{coro}
\pre

By definition of $\mathcal R$, there exists an invariant state $\rho$ with $\supp\,\rho\cap \mathcal V \neq \{0\}$. By Proposition \ref{prop_coherence}, $P_\V \,\rho\, P_\V$ is (up to normalization) an invariant state with support in $\V$.
\fin
\smallskip

The following Proposition is the main result in this section:
\begin{prop}\label{prop_minimal_enclosures}
A subspace of $\mathcal R$ is a minimal enclosure if and only if it is the support of an extremal invariant state. Moreover, any enclosure included in~$\mathcal R$ contains a (nontrivial) minimal enclosure. Equivalently, for any invariant state $\rho$, there exists an extremal invariant state $\rho_{\mathrm{ex}}$ with $\supp\,\rho_{\mathrm{ex}} \subset \supp\, \rho$.
\end{prop}

\pre

If $\mathcal V$ is a minimal enclosure contained in $\mathcal R$, then by Corollary \ref{coro_subinvariantstate}, there exists a $\Phi$-invariant state $\rho_\V$ with support in $\V$. By the discussion following Definition~\ref{defi_enclosures}, the restriction of $\Phi$ to ${\mathcal I_1(\V)}$ is irreducible. Proposition \ref{prop_Schrader} shows that $\rho_\V$ is the unique $\Phi$-invariant state with support in $\V$, and $\supp\,\rho_{\mathcal V}=\V$. This $\rho_{\mathcal V}$ must be extremal since $\rho_{\mathcal V}=t\, \rho_1 + (1-t)\, \rho_2$ with $\rho_1$, $\rho_2$ invariant states and $t\in(0,1)$ would imply that $\rho_1$, $\rho_2$ are invariant states with support in $\mathcal V$ but then by uniqueness, $\rho_{\mathcal V}=\rho_1=\rho_2$.

Conversely, if $\mathcal V=  \mathrm{supp}\, \rho$ with $\rho$ an extremal invariant state, then by Proposition \ref{prop_SuppInvSt}, $\mathcal V$ is an enclosure. If we suppose, by contradiction, that it is not minimal, then there exists an enclosure $\mathcal W$ with $\mathcal W \subsetneq\mathcal V\subset \mathcal R$ and, by Corollary~\ref{coro_subinvariantstate}, an invariant state $\rho'$ with $\supp\,\rho'\subset \mathcal W$. Since $\rho$ is faithful on $\mathcal V$, by the same argument as in the proof of Proposition \ref{prop_SuppInvSt}, 
we can approximate $\rho'$ in the $\mathcal I_1(\mathcal V)$ norm sense by a sequence $(\rho'_p)_p$ of  finite-dimensional operators such that for every $p$, there exists $\lambda_p$ with $\rho'_p\leq\lambda_p \rho$. If we let $\Psi_n=\frac 1n \sum_{k=0}^{n-1} \Phi^k$ then by a standard compacity argument, $(\Psi_n(\rho'_p))_n$ converges weakly to a $\Phi$-invariant nonnegative trace-class operator $\rhoinv_p$ which therefore satisfies $\rhoinv_p \leq \lambda_p \,\rho$. The extremality of $\rho$ implies that $\rhoinv_p$ is proportional to $\rho$. This in turn implies that $(\Psi_n(\rho'))_n$ converges weakly to $\rho$, but $\Psi_n(\rho')=\rho'$ by the $\Phi$-invariance of $\rho'$. Therefore, $\rho'=\rho$, a contradiction.

By Proposition \ref{prop_SuppInvSt} and Corollary \ref{coro_subinvariantstate}, the second claim and third claims are equivalent. To prove the second one, consider the maps $\Phi^*_{\mathcal R}$ on the set $\mathcal B(\mathcal R)$ of bounded operators acting on ${\mathcal R}$ defined by 
$$
\Phi^*_{\mathcal R}(P_{\mathcal R} x P_{\mathcal R}) = P_{\mathcal R} \Phi^*(x) P_{\mathcal R},$$
and denote by ${\mathcal F}(\Phi^*_{\mathcal R})$ the vector space of the fixed points for $\Phi^*_{\mathcal R}$, i.e.
${\mathcal F}(\Phi^*_{\mathcal R}) =\{ X\in P_{\mathcal R}{\mathcal B}(\H)P_{\mathcal R}: \Phi^*_{\mathcal R}(X)=X\}.$
We know that ${\mathcal F}(\Phi^*_{\mathcal R})$ is the image of a normal conditional expectation by Theorem 2.1 of \cite{FV}. The proof of Theorem~5 of \cite{Tom} shows then that ${\mathcal F}(\Phi^*_{\mathcal R})$ is an atomic subalgebra. 
 It is trivial to verify that the projections contained in ${\mathcal F}(\Phi^*_{\mathcal R})$ are exactly the projections on enclosures contained in ${\mathcal R}$. So, for any enclosure $\mathcal V$, we consider the corresponding projection $P_{\mathcal V}\in {\mathcal F}(\Phi^*_{\mathcal R})$; but since ${\mathcal F}(\Phi^*_{\mathcal R})$ is atomic, it contains a minimal projection $P'\le P$ and the range of $P'$ is then a minimal enclosure contained in ${\mathcal V}$. \qed

\begin{remark}
The proof of point 3 of Proposition \ref{prop_minimal_enclosures} can be given in a more constructive way: consider an invariant state $\rho$, which by restriction one can assume is faithful, i.e. with support $\H$. By the Banach-Alaoglu theorem, the set $\mathcal B(\mathcal H)^*_{+,1}\cap \mathcal F(\Phi)$ is a compact, convex, metrizable subset of the locally convex space  $\mathcal B(\mathcal H)^*$ equipped with the weak-* topology. By Theorem 4.1.11  and Proposition 4.1.3 in \cite{BR1}, and the fact that affine maps on $\mathcal B(\mathcal H)^*$ are exactly the maps $\eta \mapsto \eta(X)$ for $X \in \mathcal B(\mathcal H)$, there exists a Borel probability measure $\mu$ in~$\mathcal B(\mathcal H)^*$, such that $\rho(X)= \int\eta(X) \mathrm{d} \mu(\eta)$ for any $X$, and $\mu$ has support in the set of extremal states of $\mathcal B(\mathcal H)^*_{+,1}\cap \mathcal F(\Phi)$. Since in addition the set $\mathcal S(\H) \cap \mathcal F(\Phi)$ is a face,  $\mu$ has support in the set of extremal states of $\mathcal S(\H)\cap \mathcal F(\Phi)$. For any Borel set $B$ of $\mathcal B(\mathcal H)^*$ with $\mu(B)>0$ one can define $\rho_B=\frac1{\mu(B)}\int_B \eta(X) \mathrm{d} \mu(\eta)$. This $\rho_B$ is a state with $\supp\, \rho_B \subset \supp\, \rho$. By considering a sequence of Borel sets that are balls $B(\rho_0,\frac1n)$ for the metric compatible with the weak-* topology restricted to the unit sphere of $\mathcal B(\mathcal H)^*$, one has for $\mu$-almost all $\rho_0$ that~$\rho_{B(\rho_0,\frac1n)} \to \rho_0$ in the topology of $\mathcal S(\H)$, so that $\supp\, \rho_0 \subset \supp\,\rho$.
\end{remark}

For any quantum channel $\Phi$, point 2 of Proposition \ref{prop_coherence}, together with Proposition \ref{prop_minimal_enclosures}, will allow us to decompose the space $\mathcal R$ associated with $\Phi$ into a direct sum of minimal enclosures, and each of them is the support of an extremal invariant state. We give the following sequel to the two results quoted above, that essentially shows that the procedure of taking orthogonal complements is efficient in terms of decomposition into minimal enclosures:
\begin{lemme} \label{lemme_makeitorthogonal}
Let $\mathcal V = \V_1+\ldots+ \V_n+\V_{n+1}$, where the $\V_i$, $i=1,\ldots, n+1$, are distinct minimal enclosures contained in $\mathcal R$, and $\V_i\perp \V_j$ for $i\neq j$ in $1,\ldots, n$.  Then there exists a minimal enclosure $\V_{n+1}'$, orthogonal to $\V_1,\ldots, \V_n$ and such that $\V= \V_1+\ldots+\V_n+\V_{n+1}'$. If $n=1$ then one can take $\V_2'= \V\cap \V_1^\perp$. In particular, if a subspace of $\mathcal R$ can be written as a sum of minimal enclosures, then it can be written as a sum of mutually orthogonal minimal enclosures.
\end{lemme}
\pre
Let us first prove the claim for $n=1$. We know that $\V$ is an enclosure as direct sum of two enclosures and so by Proposition \ref{prop_coherence}, $\V_2'$ is an enclosure. If $\V_2\perp \V_1$ then $\V_2'=\V_2$ and there is nothing to prove. Assume therefore that $\V_2\not\perp \V_1$. Proposition \ref{prop_minimal_enclosures} provides us with a nontrivial minimal  enclosure ${\mathcal W}\subseteq \V_2'$. Then $\mathcal W \not\subset \V_2$ for otherwise $\mathcal W =\V_2\subset \V_2'$ and $\V_2\perp \V_1$, a contradiction. Since ${\mathcal W}$ is contained in $\V_1+\V_2$, there exists $w\in {\mathcal W}$ such that $w=v_1+v_2$ for some $v_1\in \V_1\setminus \{0\}$ and $v_2\in \V_2$. Then $v_1=w-v_2\in \V_1\cap({\mathcal W}+\V_2)$. By Corollary \ref{coro_sumsenclosures}, this means that $\V_1\cap({\mathcal W}+\V_2)$ is a nontrivial enclosure contained in the minimal enclosure $\V_1$. Consequently $\V_1\subset {\mathcal W}+\V_2$, so that $\V_1+\V_2\subset{\mathcal W}+\V_2$ and necessarily ${\mathcal W}=\V_2'$. This proves the minimality of $\V_2'$.

Now if $n>1$, define $\V_{n+1,1}'=(\V_1+\V_{n+1})\cap \V_1^\perp$. By the preceding discussion, $\V_{n+1,1}'$ is orthogonal to $\V_1$ and $\V_1+\V_{n+1} = \V_1+\V_{n+1,1}'$. Then define $\V_{n+1,2}'=(\V_2+\V_{n+1,1}')\cap \V_2^\perp$. This $\V_{n+1,2}'$ is now orthogonal to $\V_1$ and $\V_2$ and $\V_2+\V_{n+1,1}'=\V_2+\V_{n+1,2}'$ so that $\V_1+\V_2+\V_{n+1}=\V_1+\V_2+\V_{n+1,2}'$. Iterating this process gives the desired $\V_{n+1}'$ in the form of $\V_{n+1,n}'$.
\fin
\smallskip

We therefore have our main tool for decompositions of quantum channels into irreducible ones. We wish to relate these decompositions to  the structure of invariant states of $\Phi$. In the case of Markov chains, it is well-known that these are all convex combinations of the extremal invariant states associated with irreducible parts in the decomposition. We will see in the next section, however, that this is not the case for general quantum channels.

\section{Invariant states of non-irreducible quantum channels}
\label{section_decomposition}

In this section we study the last ingredient of our decomposition, that is, how the invariant states of a quantum channel on a sum $\V_1+\V_2$ of two minimal enclosures relate to the extremal invariant states associated with these two minimal enclosures. We will see that this relation will depend on the uniqueness of the decomposition $\V_1+\V_2$.

Let us define what we mean by this uniqueness. We say that the decomposition of a subspace $\mathcal Z$ of $\mathcal R$ in a direct sum of minimal enclosures is unique, if, whenever $(\mathcal V_\alpha)_{\alpha\in A}$ and $(\mathcal W_\beta)_{\beta\in B}$ are two families of minimal enclosures with
\[ \V_\alpha\cap \V_{\alpha'}=\{0\}  \mbox{ for any } \alpha\neq \alpha',\qquad  \mathcal W_\beta\cap \mathcal W_{\beta'}=\{0\} \ \mbox{ for any } \beta\neq \beta',\]
and $\mathcal Z=\sum_{\alpha\in A}\mathcal V_\alpha = \sum_{\beta\in B} \mathcal W_\beta,$
then the sets $\{\mathcal V_\alpha, \, \alpha\in A\}$ and $\{\mathcal W_\beta,\, \beta\in B\}$ coincide, and in particular $A$ and~$B$ have the same cardinality.
\smallskip

The following lemma characterizes the situations when the decomposition of a subspace as the direct sum of two enclosures is unique. First remark that, by point $2$ in Proposition \ref{prop_coherence}, if $x$ and $y$ are in $\mathcal R$ then 
\begin{itemize}
\item either $\enc(x )\perp \enc(y)$,
\item or $ x\not\in\enc(y)^\perp$ and $y\not\in\enc(x)^\perp$.
\end{itemize}
Indeed, if $y \in \enc(x)^\perp\cap \mathcal R$ then $\enc(y)\perp \enc(x)$.

\begin{lemme}\label{lemma_uniquenessdec} 
Let $\mathcal V = \V_1 + \V_2$, where $\V_1$ and $\V_2$ are minimal enclosures contained in $\mathcal R$. The decomposition of $\mathcal V$ in a direct sum of minimal enclosures is unique if and only if any enclosure $\mathcal W$ such that $\mathcal W\not\perp \V_1$ and $\mathcal W\not\perp \V_2$ satisfies $\mathcal W \cap \mathcal V = \{0\}$.  If the latter statement holds, then the two enclosures are orthogonal.
\end{lemme}

\pre

Assume the decomposition of $\mathcal V$ as a direct sum of minimal enclosures is unique. Then $\V_1 \perp \V_2$, otherwise by Proposition \ref{prop_coherence},  $\mathcal V \cap \V_1^\perp$ would be an enclosure that does not contain $\V_2$, leading to a different decomposition of $\mathcal V$. Now consider a minimal enclosure $\mathcal W$ with $\mathcal W\not\perp \V_1$ and $\mathcal W\not\perp \V_2$. This implies $\mathcal W\neq \V_1$ so by Proposition \ref{prop_enclosures}, $\mathcal W \cap \V_1=\{0\}$. If $\mathcal W\cap \mathcal V\neq \{0\}$ then it is an enclosure in $\mathcal W$ so by minimality, $\mathcal W\subset \mathcal V$. Then $\mathcal W \oplus \V_1$ is a direct sum of minimal enclosures contained in $\mathcal V$, so, by Proposition \ref{prop_minimal_enclosures}, one can complete this as a decomposition of~$\mathcal V$ into a direct sum of minimal enclosures. This is a contradiction, leading to $\mathcal W\cap \mathcal V= \{0\}$.

Now assume that any enclosure $\mathcal W$ such that $\mathcal W\not\perp \V_1$ and $\mathcal W\not\perp \V_2$ satisfies $\mathcal W \cap \mathcal V = \{0\}$. 
Taking first $\mathcal W= \V_2$, which obviously has a nontrivial intersection with $\mathcal V$, we obtain that $\V_1\perp \V_2$.
Now consider some minimal enclosure $\V_3$ contained in $\mathcal V$. Then, by assumption, one has \textit{e.g.} 
$\V_3\perp \V_1$ and $\V_3\not\perp \V_2$ and so $\V_3\subset \V_1^\perp \cap \mathcal V$, which, as proved above, is $\V_2$. This proves the uniqueness of the decomposition. \fin


Next we need to strengthen Proposition \ref{prop_coherence} to distinguish between the situations where the decomposition into minimal enclosures is unique or not. The first result treats the situation where the decomposition is unique. To simplify the notation, from now on, when $\mathcal V$ is an enclosure, we will denote by $\Phi_{|\mathcal V}$ (instead of $\Phi_{|\mathcal I_1(\mathcal V)}$) the restriction of $\Phi$ to $\mathcal I_1(\mathcal V)$.
\begin{prop} \label{prop_enclosures_unique}
If $\rho$ is $\Phi$-invariant and $\mathcal V$ and $\mathcal W$ are two minimal enclosures contained in $\mathcal R$, such that the decomposition of $\mathcal V + \mathcal W$ into a sum of minimal enclosures is unique, then $P_\V\, \rho\, P_{\mathcal W}=P_{\mathcal W}\,\rho \,P_\V=0$, \textit{i.e.} with the notation of Proposition~\ref{prop_coherence} one has $\rho_{\mathcal C}=\rho_{\mathcal C}'=0$.
\end{prop}

\pre
If $\mathcal V$ and $\mathcal W$ are minimal enclosures in $\mathcal R$, then, by Proposition \ref{prop_minimal_enclosures}, they are the supports of extremal invariant states $\rho_{\mathcal V}$ and $\rho_{\mathcal W}$. Because the decomposition of $\mathcal V+ \mathcal W$ into minimal enclosures is unique, $\rho_{\mathcal V}$ and~$\rho_{\mathcal W}$ are the unique extremal invariant states of $\Phi_{|(\mathcal V +\mathcal W)}$. Since the set of invariant states is convex, then by the Krein-Milman theorem, $\rho$ is a convex combination of $\rho_{\mathcal V}$ and $\rho_{\mathcal W}$, so $\rho_{\mathcal C}$ and $\rho_{\mathcal C}'$ must be zero. 
\fin
\begin{remark} \label{remark_MarkovChainCase}
Consider the quantum channel $\Phi$ associated with a Markov chain as in Example \ref{example_MarkovChain}. It is a simple observation that a minimal enclosure for $\Phi$ is necessarily of the form $\V=\ell^2(C)$ for $C$ a minimal communication class for the Markov chain (where $\ell^2(C)$ is viewed as a subspace of $\ell^2(E)$). Therefore, two distinct minimal enclosures $\V_1$ and $\V_2$ are necessarily orthogonal, decompositions into sums of minimal enclosures are unique, and any invariant state on $\H=\ell^2(V_1+V_2)$ is a convex combination of the extremal invariant states $\rho_1,~\rho_2$ with supports $\ell^2(V_1)$, $\ell^2(V_2)$ respectively.
\end{remark}
\smallskip

A second result will allow us to describe more explicitly the situation where the decomposition into minimal enclosures is not unique, and describe the associated invariant states:
\begin{prop}\label{prop_partialisom}
Let $\V_1$ and $\V_2$ be  two minimal enclosures contained in $\mathcal R$. Assume that the decomposition of $\mathcal V =  \V_1+ \V_2$ in a direct sum of minimal enclosures is not unique. Then  $\dim\,\V_1= \dim \,\V_2.$
If, in addition, $\V_1\perp  \V_2$ (as can be chosen by Lemma \ref{lemme_makeitorthogonal}) then there exists a partial isometry $Q$ from $\V_1$ to~$\V_2$ satisfying
\begin{equation}\label{eq_partialisometry}
Q^* Q = \id_{|\V_1}\qquad Q\,Q^* = \id_{|\V_2}
\end{equation}
and for any $\rho$ in $\mathcal I_1(\mathcal H)$, and $R=QP_{\V_1}+Q^*P_{\V_2}$:
\begin{equation}\label{eq_commpartialisom}
R\, \Phi(\rho)\, P_{\V_i} + P_{\V_i}\,  \Phi(\rho)\, R= \Phi\big(R\,\rho\, P_{\V_i} + P_{\V_i}\,\rho\, R\big) \quad \mbox{for }i=1,2.
\end{equation}
\end{prop}

\pre

By Lemma \ref{lemma_uniquenessdec}, there exists a minimal enclosure $\mathcal W$ in $\V_1+\V_2$ such that $\mathcal W \not\perp \V_i$, $i=1,2$. Then \textit{e.g.} $\V_1\cap \mathcal W^\perp$ is a nontrivial enclosure contained in $\V_1$, and by minimality $\V_1\subset \mathcal W^\perp$. Therefore $\dim \V_1 \leq \dim\mathcal W$, and by symmetry one has the equality $\dim\,\V_1 = \dim\,\mathcal W$. Similarly one has $\dim\,\V_2 = \dim\,\mathcal W$.

Assume now that $\V_1\perp \V_2$. Define the map  $\Phi_{\mathcal R}^*$ as in the proof of Proposition~\ref{prop_minimal_enclosures}. By Remark \ref{remark_PFHeisenberg}, if $E=\V_1$, $\V_2$ or $\mathcal W$, then $P_{E}$ is (up to multiplication) the unique invariant of the restriction $\Phi_{E}^*$ of $\Phi_{\mathcal R}^*$ to $\mathcal B(E)$. Consider the decomposition of $P_{\mathcal W}=\begin{pmatrix}A& B^*\\B&C\end{pmatrix}$ in the splitting $\mathcal V=\V_1\oplus \V_2$, where necessarily~$B\neq 0$. A simple consequence of Proposition \ref{prop_coherence} is that in the same decomposition, 
$\Phi_{\mathcal R}^*(P_{\mathcal W})=\begin{pmatrix}{\Phi_{\mathcal R}^*}(A)& {\Phi_{\mathcal R}^*}(B)^*\\ {\Phi_{\mathcal R}^*}(B)&{\Phi_{\mathcal R}^*}(C)\end{pmatrix}$. 
Therefore $A$ is proportional to $P_{\V_1}$ and $C$ to $P_{\V_2}$. Writing relations $P=P^*=P^2$ satisfied by $P_{\mathcal W}$, one sees that $B$ must be proportional to an operator $Q$ satisfying relations~\eqref{eq_partialisometry}. Fix $Q$; for $\theta\in[0,\pi]$, the operator defined by
\[P_{\theta}=\begin{pmatrix}\cos^2 \theta & \sin \theta\cos\theta\, Q^*\\ \sin \theta\cos\theta\, Q & \sin^2\theta \end{pmatrix} \]
is an orthogonal projection preserved by the map $\Phi_{\mathcal R}^*$. So its range is an enclosure and, by point $3$ of Proposition \ref{prop_coherence}, $P_\theta$ will satisfy the relation
\[\Phi(P_{\theta} \, \rho \, P_{\theta})= P_{\theta} \,\Phi(\rho)\,P_{\theta},\]
for any $\rho$ in $\mathcal I_1(\mathcal H)$.
Differentiating this relation with respect to $\theta$, we have
\[
\Phi\big(\frac{\mathrm d P_\theta}{\mathrm d\theta} \, \rho \, P_{\theta}+ P_{\theta} \, \rho \, \frac{\mathrm d P_\theta}{\mathrm d\theta}\big)=\frac{\mathrm d P_\theta}{\mathrm d\theta} \, \Phi(\rho) \, P_{\theta}+ P_{\theta} \, \Phi(\rho) \, \frac{\mathrm d P_\theta}{\mathrm d\theta}.
\] 
Computing the derivatives
at $\theta=0$ and $\theta=\pi/2$, we obtain relations \eqref{eq_commpartialisom}.
\fin

\begin{coro}\label{coro_partialisom2}
Assume that $\mathcal V = \V_1+ \V_2$ where $\V_1$ and $\V_2$ are mutually orthogonal minimal enclosures, contained in $\mathcal R$, but that the decomposition of $\V$ into a direct sum of minimal enclosures is non-unique. For  $i=1,2$ let $\rhoinv_i$ be the unique invariant state with support in $\V_i$, if it exists, and $\rhoinv_i=0$ otherwise. Consider $Q$ the partial isometry defined in Proposition \ref{prop_partialisom}. Then $\rhoinv_2=Q\, \rhoinv_1 \, Q^*$.

If $\rho$ is an invariant state with support in $\mathcal V$, then:
\begin{itemize}
\item $P_{\V_1}\,\rho\,P_{\V_1} $ is proportional to $\rhoinv_1$,
\item $P_{\V_2}\,\rho\,P_{\V_2}$ is proportional to $\rhoinv_2$,
\item $P_{\V_1}\,\rho\,P_{\V_2}$ is proportional to $\rhoinv_1\, Q^*=Q^*\rhoinv_2$,
\item $P_{\V_2}\,\rho\,P_{\V_1}$ is proportional to $\rhoinv_2\, Q=Q\rhoinv_1$.
\end{itemize}
\end{coro}

\pre
The first identity is obtained by applying relation \eqref{eq_commpartialisom} to $\rho=\rhoinv_1$ with~$P_1$, then applying it again to the resulting relation, this time with $P_2$.

That each $\rho_{i,j}=P_{\V_i}\rho P_{\V_j}$ is an invariant is an immediate consequence of Proposition \ref{prop_coherence}. The relation satisfied by $\rho_{1,2}$ and $\rho_{2,1}$ is then obtained by applying relation~\eqref{eq_commpartialisom} to e.g. $\rho_{1,2}$, with $P_1$ or $P_2$.
\fin

\section{Irreducible decompositions of quantum channels and invariant states} \label{section_irreducibledecompositions}
We are now in a position to state the relevant decomposition associated with $\Phi$. 

\begin{prop}\label{prop_finaldec}
Let $\Phi$ be a quantum channel on a separable Hilbert space $\H$. There exists a decomposition of $\H$ in the form
\begin{equation}\label{eq_finaldec}
\H = \mathcal D + \sum_{\alpha \in A}\V_\alpha + \sum_{\beta \in B}\sum_{\gamma \in C_\beta} \V_{\beta,\gamma},
\end{equation}
where any set $A,B,C_\beta$ is at most countable, $A$ and $B$ can be empty (but
not simultaneously), any $C_\beta$ has cardinality at least two, and:
\begin{itemize}
\item every $\V_\alpha$ or $\V_{\beta,\gamma}$ in this decomposition is a minimal enclosure,
\item for $\beta$ in $B$, any minimal enclosure that is not orthogonal to $\sum_{\gamma \in C_\beta}\V_{\beta,\gamma}$ is contained in $\sum_{\gamma \in C_\beta} \V_{\beta,\gamma}$,
\item any two distinct subspaces $\mathcal D$, $\V_\alpha$, $\V_{\beta,\gamma}$ are mutually orthogonal.
\end{itemize}
\end{prop}

\pre

We start with the orthogonal decomposition $\mathcal H = \mathcal D +\mathcal R$, and proceed to decompose $\mathcal R$. Consider the set of all minimal enclosures $\V$ with the property that any minimal enclosure different from $\V$ is orthogonal to $\V$. By separability, this set is at most countable. Then we can denote all such minimal enclosures by $\V_\alpha$, with $\alpha$ in a (countable) set of indices $A$. Let $\mathcal O$ be the direct sum of all these enclosures, $\mathcal O=\sum_{\alpha\in A}\V_\alpha$. 
Then $\mathcal O$ is an enclosure, and, by point $2$ of Proposition \ref{prop_coherence}, $\mathcal R \cap \mathcal O^\perp$ is also an enclosure.

Assume that $\mathcal R \cap \mathcal O^\perp$ is nontrivial; we proceed to decompose it. Let $\beta(1)=1$ and consider a minimal enclosure $\V_{\beta(1),1}\subset \mathcal R \cap \mathcal O^\perp$. By the definition of $\mathcal O$, there exists a minimal enclosure $\V_2$  in $\mathcal R \cap \mathcal O^\perp$, and by Lemma \ref{lemme_makeitorthogonal} we can choose $\V_{\beta(1),2}$ minimal, orthogonal to $\V_1$, and such that $\V_{\beta(1),1}+\V_{\beta(1),2}=\V_{\beta(1),1}+\V_2$. If all minimal enclosures are either included in $\V_{\beta(1),1}+ \V_{\beta(1),2}$ or orthogonal to $\V_{\beta(1),1}+ \V_{\beta(1),2}$, we set $C_{\beta(1)}=\{1,2\}$. Otherwise, we call $\V_{3}$ a minimal enclosure not included in and not orthogonal to $\V_{\beta(1),1}+\V_{\beta(1),2}$. By Lemma \ref{lemme_makeitorthogonal} we can choose  $\V_{\beta(1),3}$ minimal, orthogonal to $\V_{\beta(1),1}+\V_{\beta(1),2}$ and such that $$\V_{\beta(1),1}+\V_{\beta(1),2}+\V_{\beta(1),3}=\V_{\beta(1),1}+\V_{\beta(1),2}+\V_{3}$$
and we proceed again with the same method for a denumerable number of steps so that we construct $C_{\beta(1)}$.
If $\mathcal R\cap \mathcal O^\perp \cap \big(\sum_{\gamma\in C_{\beta(1)}}E_{\beta(1),\gamma}\big)^\perp\neq \{0\}$, we can iterate the procedure.
\fin
\medskip

Before we state our next result, let us give some notation. We fix a decomposition \eqref{eq_finaldec} as considered in Proposition \ref{prop_finaldec}. We define
\[P_0 = P_{\mathcal R^\perp},\qquad 
P_i = P_{\V_i}\;  \mbox{ for $i\in A$ or $i\in \bigcup_{\beta\in B}\{\beta\}\times C_\beta$},
\]
and, for a state $\rho$, and $i$, $j$ taking the values $0$, $\alpha \in A$ or $(\beta,\gamma)\in \bigcup_{\beta\in B}\, \{\beta\}\times C_\beta$
\begin{equation}
\label{eq_decomprho} \rho_i=P_i \, \rho \, P_i\qquad  \rho_{i,j}= P_i \, \rho \, P_j.
\end{equation}
In addition, we denote by $\rhoinv_i$ the unique invariant state of $\Phi_{|\V_i}$ if it exists, and $\rhoinv_i=0$ otherwise.
\smallskip

We can now state:
\begin{theo}\label{theo_invariantstates}
Let $\rho$ be a $\Phi$-invariant state and consider a related orthogonal decomposition of the form \eqref{eq_finaldec}. With the notation~\eqref{eq_decomprho}, we have
\begin{enumerate}
\item $\rho_0=0$,
\item every $\rho_i$ is proportional to $\rhoinv_i$, for all indices $i\in A\cup\bigcup_{\beta\in B}\{\beta\}\times C_\beta$,
\item for $\gamma\neq \gamma'$ in $C_\beta$, the off-diagonal term $\rho_{((\beta,\gamma),(\beta,\gamma'))}$, which we simply denote by $\rho_{(\beta,\gamma,\gamma')}$, may be non-zero, and is $\Phi$-invariant. In addition, there exists a partial isometry $Q_{(\beta,\gamma,\gamma')}$ from $\V_{\beta,\gamma}$ to $\V_{\beta,\gamma'}$ such that:
\begin{itemize}
\item $\rhoinv_{(\beta,\gamma')}=Q_{(\beta,\gamma,\gamma')}\, \rhoinv_{(\beta,\gamma)}\, Q_{(\beta,\gamma,\gamma')}^*$
\item $\rho_{(\beta,\gamma,\gamma')}$ is proportional to $Q^*_{(\beta,\gamma,\gamma')}\,\rhoinv_{(\beta,\gamma')}=\rhoinv_{(\beta,\gamma)}\, Q^*_{(\beta,\gamma,\gamma')}$,
\end{itemize}

\item all other $\rho_{i,j}$ (for $i,j$ taking all possible values in $\{0\}\cup A\cup\bigcup_{\beta\in B}\{\beta\}\times C_\beta$) are zero.
\end{enumerate}
\end{theo}

\pre

This follows from a repeated application of Propositions \ref{prop_coherence} and \ref{prop_finaldec}, and Corollary \ref{coro_partialisom2}. \fin

\begin{remark}
The decomposition of an invariant state $\rho$ given by Theorem \ref{theo_invariantstates} can be rewritten in the same form as in formula $(12)$ of Theorem $7$ in \cite{BN}, or as in Theorem 22 of \cite{DFSU}, by simple algebraic manipulations. 
The key object is an isomorphism between $\V_{\beta,1}\otimes \cc^{C_\beta}$ and $\sum_{\gamma \in C_\beta} \V_{\beta,\gamma}$ for each $\beta$, given by
\[\mathcal E (u\otimes x) = \sum_{\gamma \in C_\beta} u_\gamma \, Q_{(\beta,1,\gamma)} x \ \mbox{ for }\ u=(u_\gamma)_{\gamma\in\C_\beta}.\]
\end{remark}

\begin{remark}
The representation of invariant states appearing in Theorem \ref{theo_invariantstates} has recently been studied in \cite{DFSU}, where an analogous result is proven in infinite dimension (and in the continuous time setting, but this point is not crucial). Our techniques and starting points are completely different and essentially replicate the approach used in \cite{BN} and \cite{CP1}.
Concerning the orthogonal decomposition and the representation of invariant states, however, our result is more general than the one in \cite[Theorem 2.1]{DFSU}, since we do not need to assume the atomicity of the decoherence free algebra (notice that the existence of a faithful normal invariant state assumed in \cite{DFSU} is not a restriction, since our decomposition is anyway only for the fast recurrent subspace $\mathcal R$, and by Remark \ref{remark_Umanita}, the restriction of~$\Phi$ to~$\mathcal R$ has a faithful invariant state). The key step which allows us to avoid this additional assumption is that we can prove that the fixed point algebra~${\mathcal F}(\Phi^*_{\mathcal R})$ is atomic. When there exists a faithful invariant state, this means that ${\mathcal F}(\Phi^*)$ is atomic. However, we do not know whether the decoherence free algebra (see \cite{DFSU}), usually denoted by ${\mathcal N}(\Phi^*)$,  is atomic, neither can we so far deduce other generalizations of the results on the structure of this algebra studied in \cite{DFSU}. \end{remark}

%
%
%
%
%
%
%
%
\section{Examples}\label{section_examples}

\begin{example} (classical Markov chains)
Consider as in Example \ref{example_MarkovChain} a Markov chain on a countable set $E$. Denote by $(C_\alpha)_{\alpha\in A}$ the family of minimal communication classes $C_\alpha$ such that the Markov chain has an invariant probability~$\pi^{(\alpha)}$ with support $C_\alpha$, by $R=\cup_{\alpha\in A} C_\alpha$ the (disjoint) union of these classes, and by~$D$ the complement $D=E\setminus R$. Then, according to the discussion in Remark \ref{remark_MarkovChainCase}, the decomposition \eqref{eq_finaldec} of $\H=\ell^2(E)$ is given by 
\[ \H = \mathcal D + \sum_{\alpha\in A} \V_\alpha \quad \mbox{where}\ \mathcal D = \ell^2(D), \ \V_\alpha = \ell^2(C_\alpha)\]
and any invariant state on $\H$ is a convex combination of the extremal states, which are of the form $\sum_{i\in C_\alpha} \pi^{(\alpha)}_i \ketbra{e_i}{e_i}$.
\end{example}

\begin{example}
Consider the quantum channel defined in Example \ref{example_23}.
From the computations in Example \ref{example_23}, one has $\mathcal R=\cc\, e_1$ and therefore $\mathcal D = \cc\, e_2$.
\end{example}

\begin{example} ($2\times 2$ matrices)
Consider $\H=\cc^2$ and $\Phi$ a positive quantum map on the algebra $\mathcal B(\cc^2)$, which we identify with the set $M_2(\cc)$ of $2\times 2$ matrices and equip with the scalar product $ \langle x , y \rangle_{M_2} = \mathrm{tr} (x^*y)$. The Pauli matrices
\[
\sigma_0=\frac{1}{\sqrt{2\,}}\, \id_{\cc^2}, \quad
   \sigma_1 =\frac{1}{\sqrt{2\,}}\begin{pmatrix}
      0 & 1 \\ 1 & 0
    \end{pmatrix} \quad
\sigma_2 = \frac{1}{\sqrt{2\,}}\begin{pmatrix}
      0 & \llap{-}i \\ i & 0
    \end{pmatrix} \quad
\sigma_3 = \frac{1}{\sqrt{2\,}}\begin{pmatrix}
       1 & 0 \\ 0 &  \llap{-}1
  \end{pmatrix}
\]
form an orthonormal basis of $M_2(\cc)$ and satisfy
\[
\sigma_k^2=\sigma_0^2,\qquad  
\sigma_k\sigma_j = - \sigma_j\sigma_k, \qquad
\sigma_j\sigma_k = i \sigma_\ell
\]
if $(j,k,\ell)\in\{(1,2,3),(2,3,1),(3,1,2)\}$. 

It is easy to see that, since $\Phi$ is trace preserving and positive, 
its matrix in the basis $\{\sigma_0,\sigma_1,\sigma_2,\sigma_3\}$ is of the form
\begin{eqnarray}\label{generator}
\Phi=
\begin{pmatrix}
1 & {}^t 0\\
b & A
\end{pmatrix}
\end{eqnarray}
where $b\in {\mathbb R}^3$, ${}^t 0 = (0,0,0)$, $A$ is a $3\times 3$ matrix with real coefficients.
The map~$\Phi$ is positive if and only if $\|b+Ax\|\le 1$ for all $x$ such that $\|x\|\le 1$
(see \cite{C} for more details, even if in the continuous time setting).
\smallskip

It is well-known that states on $\cc^2$ are all operators of the form $\rho=\sigma_0+u\cdot\sigma$ with $u$ in ${\mathbb R}^3$, $\|u\|\leq 1$ (here we use the standard notation $u\cdot\sigma= \sum_{i=1,2,3}u_i\, \sigma_i$). This is called the Bloch sphere representation.
In addition, it is easy to see that a state $\rho=\sigma_0+u\cdot\sigma$ is invariant for $\Phi$ if and only if $b+Au-u=0$.

Essentially, the problem of decomposing $\cal R$ into minimal enclosures is reduced to solving the linear system $b+Au-u=0$, and then considering if there exist solutions with $\|u\|=1$ and how many they are.
However, by the Markov-Kakutani Theorem, an invariant state always exists.
For the decomposition of the fast recurrent space $\cal R$, only $3$ different cases are possible.
\begin{itemize}
\item There exists a unique invariant state $\rho$. Then the only minimal enclosure in $\cal R$ is $\cal R$ itself and it has dimension $2$ when $\rho$ is faithful and $1$ otherwise.
\item There exist infinitely many invariant states, and they are given by all convex combinations of two extremal invariant states $\rho_1$ and $\rho_2$. Then $\cal R$ can be written in a unique way as the direct sum of two minimal enclosures, which are the supports of $\rho_1$ and $\rho_2$.
\item There exist infinitely many extremal invariant states. Then any state is invariant, any one dimensional subspace is an enclosure, and  $\cal R$ can be written as ${\cal R}=\cc\, e_1\oplus \cc\, e_2$, for any two linearly independent vectors $e_1$, $e_2$ in $\cc^2$. This third case is possible if and only if $\Phi$ is the identity operator.
\end{itemize} 
\end{example}

\begin{example} 
Define $V=\cc\cup\{0\}$, $\h={\mathbb C}^3$, $\H = \h\otimes \ell^2(V)$ 
and fix a canonical basis $(e_k)_{k=1,2,3}$ of $\h$ so that we represent matrices and vectors in this basis.
Choose $p,q>0$ such that $p<1/2$, $p+q<1$ and a family of operators $(L_{i,j})_{i,j \in V}$ on $\h$ such that $L_{ij}=0$ when $|i-j|\ge 2$, $L_{00}=\sqrt{1-p\,}\, \id_\h$, $L_{j+1,j}= \sqrt{p\,} \,\id_\h$ for $j\ge0$ and
\begin{eqnarray*}
L_{j-1,j}&=& \begin{pmatrix} \sqrt{1-p} &0&0\\ 0&\sqrt{1-p\,}&0 \\ 0&0&\ \sqrt{q}\  \end{pmatrix} \quad \mbox{for } j\ge1 \\
L_{j,j}&=&\sqrt{\frac{(1-p-q)}2\,}
\begin{pmatrix} 0 &0&1\\ 0&0&1\\ 0&0&0 \end{pmatrix}
\quad \mbox{for } j\ge1
\end{eqnarray*}
We have $\sum_{i\in V}L_{i,j}^*L_{ij}=\id$ for all $j$ in $V$, so that the map $\Phi$ acting on ${\mathcal I}_1(\mathcal H)$ defined by 
$$
\Phi(\rho) = \sum_{i,j\in V} (L_{i,j}\otimes \ketbra ij) \, \rho \, (L_{i,j}^*\otimes \ketbra ji),
$$
is a quantum channel.
This map $\Phi$ is an open quantum random walk with transition operators $(L_{i,j})_{i,j \in V}$ as defined in  \cite{APSS}. Denote by $(\ket j)_{j\in V}$ the canonical basis of $\ell^2(V)$. It was proved in \cite{CP1} that minimal enclosures for open quantum random walks are generated by vectors of the form $u\otimes \ket i$. Consider therefore $u={}^t(u_1,u_2,u_3)$ in $\h$, then
$$
\enc(u\otimes \ket i)
=\left\{\begin{array}{ll}
\mathrm{span}\{u\otimes \ket j, j\ge 0\} & \mbox{if } u_3=0,\\
\mathrm{span}\{e_3\otimes \ket j,(e_1+e_2)\otimes \ket j, j\ge 0\} & \mbox{if } u_3\neq0, u_1=u_2\\
{\mathcal H} & \mbox{if } u_3\neq0, u_1\neq u_2,\\
\end{array}\right.
$$
The enclosures described in the first case ($u_3=0$) are the minimal ones and so they support the extremal invariant states of the evolution. Using finite difference equations as for similar classical Markov chains, one can compute these extremal invariant states,
$$
\rho(u)=c\,\sum_{j\ge 0} \left(\frac{p}{1-p}\right)^j \ketbra uu \otimes \ketbra jj,
$$
for $u={}^t(u_1,u_2,0)\neq 0$ and a normalizing constant $c$.

Then we have $\mathcal R=\mbox{span }\{e_1,e_2\}\otimes \ell^2(V)$, $\mathcal D=\mbox{span }\{e_3\}\otimes \ell^2(V)$ and the decomposition \eqref{eq_finaldec} can be written with $A$ empty, $B$ consisting of only one element~$\beta$, $C_{\beta}=\{1,2\}$,  
$$
\V_{\beta,1} = \mathrm{span}\{v^1\otimes\ket j, j\ge 0\},
\qquad
\V_{\beta,2} = \mathrm{span}\{v^2\otimes \ket j, j\ge 0\} , 
$$
for any linearly independent vectors $v^1$ and $v^2$ orthogonal to $e_3$.
We observe that $\rhoinv_{\beta,1}=\rho(e_1)$ is the only invariant state with support $E=\mathrm{span}\{e_1\otimes \ket j, j\ge 0\}$ and, defining $Q$ as $\ketbra {e_2}{e_1}$, that any invariant state has a decomposition
$$ \rho=\begin{pmatrix} \quad  t\, \rho(e_1) \quad & \lambda \,\rho(e_1) Q^* \\ \bar\lambda \,Q\rho(e_1) & (1-t)\,   Q\rho(e_1)Q^*\end{pmatrix}
\qquad\mbox{with } t\in [0,1].
$$ 
Using the previous expressions for enclosures, one can also deduce the communication classes, in particular for the vectors of the form $u\otimes \ket j $, which are the most interesting in the special case of open quantum random walks.  
\end{example}


\bibliography{biblio}
\end{document}